\documentclass[10pt]{article}
\usepackage{latexsym}
\usepackage{amssymb}
\usepackage{amsmath}
\usepackage{amscd}
\usepackage{amsthm}
\usepackage[left=2cm,top=2.5cm,right=2.5cm,bottom=1.5cm]{geometry}
\usepackage[dvips]{graphicx}
\usepackage{epstopdf}
\usepackage{hyperref}
\begin{document}
\begin{center}
\large{\bf{An Interacting Scenario for Dark Energy in Bianchi Type-I Universe}} \\
\vspace{10mm}
\normalsize{Hassan Amirhashchi$^1$, S. N. A. Qazi$^2$ and Hishamuddin Zainuddin$^3$}\\
\vspace{5mm} \normalsize{$^1$Department of Physics, Mahshahr Branch, Islamic Azad University, Mahshahr, Iran \\
$^1$E-mail:h.amirhashchi@mhriau.ac.ir}  \\
\vspace{5mm} \normalsize{$^{2,3}$Laboratory of Computational Sciences and Mathematical Physics, Institute for
Mathematical Research, University Putra Malaysia, 43400 UPM, Serdang, Selangor D.E., Malaysia \\
$^2$E-mail: nasrullah102@gmail.com~~ $^3$E-mail: hisham@putra.upm.edu.my} \\
\end{center}
\vspace{10mm}
\begin{abstract}
We study the interaction between dark energy (DE) and dark matter (DM) in the scope of anisotropic bianchi type I space-time. First we derive the general form of the dark energy equation of state parameter (EoS) in both non-interacting and interacting cases and then we examine it's future by applying a hyperbolic scale factor. It is shown that in non-interacting case, depending on the value of the anisotropy parameter $K$, the dark energy EoS parameter is varying from phantom to quintessence whereas in interacting case EoS parameter vary in quintessence region. However, in both cases the dark energy EoS parameter $\omega^{de}$, ultimately (i. e at $z=-1$) tends to the cosmological constant ($\omega^{de}=-1$). Moreover, we fixed the cosmological bound on
the anisotropy parameter $K$ by using the recent observational data of Hubble parameter.
\end{abstract}
\section{Introduction}           
\label{sect:intro}

The direct observations and evidences made by the High-z Supernova Search Teams (Riess et al. 1998; Perlmutter et al. 1999) in 1998 and 1999 indicated that the rate of the expansion of our universe is positive i.e we live in an accelerating expanding universe. The above fact has also been approved by astrophysical observations such as measurements of cosmic microwave background (de Bernardis et al. 2000; Benoit et al. 2003; Spergel et al. 2003) and the galaxy power spectrum (Tegmark et al. 2004; Page et al. 2003). These observations show that the geometry of the present day universe is almost flat. But the most surprising and unbelievable result comes from these observations is the fact that only $\sim 4.6\%$ of the universe total energy density is in the baryonic (non-relativistic) matter, $\sim24 \%$ non-baryonic (relativistic) matter called dark matter (DM), and almost $\sim 71.4\%$ is an completely unknown component with negative pressure called dark energy (DE). Despite the matter, dark energy produces a repulsive force which give raises to the current cosmic accelerating expansion.
Since 1998, many theoretical and observational attempts have been done in order to investigate the real nature of the dark energy. The most important problem in the study of DE is the fact that this mysterious component does not interact with baryonic matter and hence we do not have any way to detect it. Although some of current observations (Bertolami et al. 2007; Le Delliou et al. 2007) show that there is an interaction between DE and DM, bout the amount of this interaction is very small and it is not detectable by our today technology. Up to date, we only know that dark energy is non-clustering and spatially homogeneous; while it dominates the the present universe, it was small at the early times.\\

From theoretical point of view, the study of the nature of dark energy is possible through it's equation of state parameter $\omega^{de}$ which is the ratio of the pressure and the energy density of DE. However, the exact value of the dark energy EoS parameter at the present time is not clear for us yet. The lack of our knowledge, allow us to suggest different theoretical candidates for dark energy. The first and natural candidate is a cosmological constant $\Lambda$ with $\omega=-1$ (Weinberg 1989; Carroll 2001). But this model can not explain why the present amount of the dark
energy is so small compared with the fundamental scale (fine-tuning problem) and why it is comparable with the critical density today (coincidence problem). To solve such fundamental problems associate with the cosmological constant scenario the different forms of dynamically changing DE with an effective equation of state EoS including quintessence ($-1<\omega^{de}<-\frac{1}{3}$) (Wetterich. 1988; Ratra \& Peebles. 1988), phantom ($\omega^{de}<-1$) (Caldwell 2002), quintom ($\omega^{de}<-\frac{1}{3}$) (Feng et al. 2005), Chaplygin gas  models (Sirvastava 2005; Bertolami et al. 2004), and etc have been proposed.\\

A simple and straightforward way to solve the coincidence problem in cosmology is to consider an energy flow from DE to DM (Cimento et al. 2003; Dalal et al. 2001). Such an energy transfer could easily explain why, at the present time, the energy densities of dark energy and dark matter are almost equal. Theoretically, interacting and non-interacting dark energy models have been widely studied in the literatures (Zhang 2005; Zimdahl \& Pavon 2004; Setare 2007a,b; Setare et al. 2009; Sheykhi \& Setare 2011; Pradhan et al. 2011a,b; Amirhashchi et al. 2011a,b,c,d ). Recently, Saha et al (2012), Saha (2013a,b), Pradhan (2013), Yadav (2012), and Yadav \& Sharma (2013) have have investigated dark energy in different contexts.\\
In this paper, we study the interaction between dark energy and dark matter on the bases of anisotropic Bianchi type I space-time. Up to our knowledge, this work is the first study of interacting dark energy in an anisotropic space-time in it's general form. The outline of the paper is
as follows: In Sect. 2, the metric and the field equations as well as the Friedmann like equation are described. Section 3 deals with the non-interacting two fluid dark energy case. The interaction between dark energy and dark matter will be studied
in Sect. 4. In Sec. 5, we constraint the anisotropy parameter $K$ using a direct fitting procedure involving the Hubble rate $H(z)$. Finally, conclusions are summarized in the last Sect. 6.

\section{The Metric and Field  Equations}
\label{sect:The Metric and Field  Equations}

In an orthogonal form, the Bianchi type I line-element is given by
\begin{equation}
\label{eq1}
ds^{2} = -dt^{2} + A^{2}(t)dx^{2}+B^{2}(t)dy^{2}+C^{2}(t)dz^{2},
\end{equation}
where $A(t), B(t)$ and $C(t)$ are functions of time only. \\

The Einstein's field equations ( in gravitational units $8\pi G = c = 1 $) read as
\begin{equation}
\label{eq2} R^{i}_{j} - \frac{1}{2} R g^{i}_{j} = T^{(m) i}_{j} +
T^{(de) i}_{j},
\end{equation}
where $T^{(m) i}_{j}$ and $T^{(de) i}_{j}$ are the energy momentum tensors of dark matter and viscous dark energy,
respectively. These are given by
\[
  T^{m i}_{j} = \mbox{diag}[-\rho^{m}, p^{m}, p^{m}, p^{m}],
\]
\begin{equation}
\label{eq3} ~ ~ ~ ~ ~ ~ ~ ~  = \mbox{diag}[-1, \omega^{m}, \omega^{m}, \omega^{m}]\rho^{m},
\end{equation}
and
\[
 T^{de i}_{j} = \mbox{diag}[-\rho^{de}, p^{de}, p^{de}, p^{de}],
\]
\begin{equation}
\label{eq4} ~ ~ ~ ~ ~ ~ ~ ~ ~ ~ ~ ~ ~ ~ = \mbox{diag}[-1, \omega^{de}, \omega^{de},
\omega^{de}]\rho^{de},
\end{equation}
where $\rho^{m}$ and $p^{m}$ are the energy density and pressure of the perfect fluid
component while $\omega^{m} = p^{m}/\rho^{m}$ is its EoS parameter. Similarly,
$\rho^{de}$ and $p^{de}$ are, respectively the energy density and pressure of the viscous DE component while
$\omega^{de}= p^{de}/\rho^{de}$ is the corresponding EoS parameter.\\

In a co-moving coordinate system ($u^{i} = \delta^{i}_{0}$),
Einstein's field equations (\ref{eq2}) with (\ref{eq3}) and
(\ref{eq4}) for Bianchi type-I metric (\ref{eq1}) subsequently
lead to the following system of equations:
\begin{equation}
\label{eq5} \frac{\ddot{B}}{B}+\frac{\ddot{C}}{C}+\frac{\dot{B}\dot{C}}{BC}=-\omega^{m}\rho^{m}-\omega^{de}\rho^{de},
\end{equation}
\begin{equation}
\label{eq6} \frac{\ddot{A}}{A}+\frac{\ddot{C}}{C}+\frac{\dot{A}\dot{C}}{AC}=-\omega^{m}\rho^{m}-\omega^{de}\rho^{de},
\end{equation}
\begin{equation}
\label{eq7}\frac{\ddot{A}}{A}+\frac{\ddot{B}}{B}+\frac{\dot{A}\dot{B}}{AB}=-\omega^{m}\rho^{m}-\omega^{de}\rho^{de},
\end{equation}
\begin{equation}
\label{eq8} \frac{\dot{A}\dot{B}}{AB}+\frac{\dot{A}\dot{C}}{AC}+\frac{\dot{B}\dot{C}}{BC}=\rho^{m}+\rho^{de}.
\end{equation}
A solution to the above set of differential equations (eqs. (\ref{eq5}-(\ref{eq8})) has been already given Saha (2005) as
\begin{equation}
\label{eq9} A(t)=a_{1}a~ exp(b_{1}\int a^{-3}dt),
\end{equation}
\begin{equation}
\label{eq10} B(t)=a_{2}a~ exp(b_{2}\int a^{-3}dt),
\end{equation}
and
\begin{equation}
\label{eq11} C(t)=a_{3}a~ exp(b_{3}\int a^{-3}dt),
\end{equation}
where
\[
a_{1}a_{2}a_{3}=1,~~~~~~~b_{1}+b_{2}+b_{3}=0.
\]
Here $a=(ABC)^{\frac{1}{3}}$ is the average scale factor of Bianchi type I model. Using eqs. (\ref{eq9})-(\ref{eq11}) in eq. (\ref{eq8}) we obtain
\begin{equation}
\label{eq12} H^{2}=\left(\frac{\dot{a}}{a}\right)^{2}=\frac{\rho^{m}+\rho^{de}}{3}+K a^{-6},
\end{equation}
which is the analogue of the Friedmann equation and $K=b_{1}b_{2}+b_{1}b_{3}+b_{2}b_{3}$, is a constant. Note that $K$ denotes the
deviation from isotropy e.g. $K=0$ represents flat FLRW universe.\\

\section{Non-Interacting Dark Energy}
\label{sect:Non-Interacting Dark Energy}

In this section we assume that there is no any interaction between dark energy (DE) and dark mater (DM). In this case, we can simply re-write the law of energy-conservation equation ($T^{ij}_{;j} = 0$) which yields
\begin{equation}
\label{eq13} \dot{\rho}^{m} + 3\frac{\dot{a}}{a}(1 + \omega^{m})\rho^{m} +
\dot{\rho}^{de} +3\frac{\dot{a}}{a}(1 + \omega^{de})\rho^{de}= 0,
\end{equation}
for these two dark components separately as
\begin{equation}
\label{eq14}\dot{\rho}^{m} + 3\frac{\dot{a}}{a}(1 + \omega^{m})\rho^{m}=0,
\end{equation}
\begin{equation}
\label{eq15}\dot{\rho}^{de}+ 3\frac{\dot{a}}{a}(1 + \omega^{de})\rho^{de}=0.
\end{equation}
integrating eq. (\ref{eq14}), we find
\begin{equation}
\label{eq16}\rho^{m} = \rho_{0}^{m} a^{-3(1+\omega^{m})},
\end{equation}
where $\rho_{0}^{m}$ is an integrating constant. \\
Using eq. (\ref{eq16}) in eq. (\ref{eq12}), we can find the energy density of the DE in terms of the average scale factor $a$ as
\begin{equation}
\label{eq17}\rho^{de} = 3H^{2}(1-\Omega^{m}_{0}a^{-3(1+\omega^{m})})-3Ka^{-6},
\end{equation}
where $\Omega^{m}=\frac{\rho^{m}}{3H^{2}}$ is the energy density of the dark mater and the subscript $0$ shows the present value of $\Omega^{m}$.\\
Now, using eqs. (\ref{eq10}), (\ref{eq11}), (\ref{eq16}), and (\ref{eq17}) in eq. (\ref{eq5}), we finally obtain the equation of state parameter (EoS) of the dark energy as
\begin{equation}
\label{eq18}\omega^{de} = \frac{2H^{2}(q-1/2)+Ka^{-6}}{3H^{2}(1-\Omega^{m}_{0}a^{-3(1+\omega^{m})})-3Ka^{-6}},
\end{equation}
\begin{figure}
\centering
\includegraphics[width=10cm,height=10cm,angle=0]{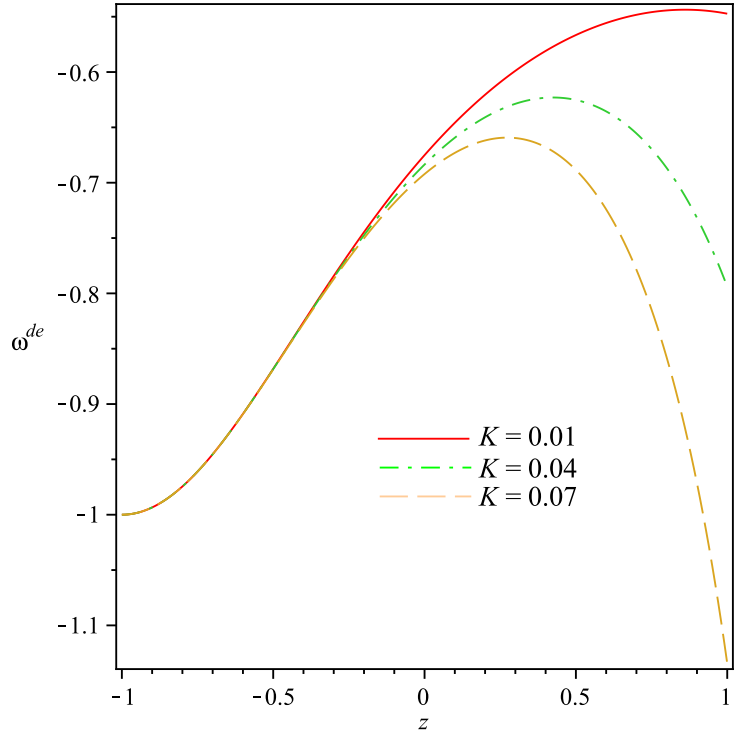}
\caption{Demo1: The EoS parameter $\omega^{de}$ versus $z$ for $K=0.01, 0.04, 0.07$, and $\Omega^{m}_{0}=0.24$.}
\label{Fig1}
\end{figure}
where $q=-\frac{\ddot{a}}{aH^{2}}$ is the deceleration parameter. This is the general form of the EoS parameter of the dark energy in the non-interacting scenario. To get more results about the behavior of the EoS parameter given by eq. (\ref{eq18}), specialty at the late time, we consider the following hyperbola scale factor
\begin{equation}
\label{eq19}a=(1+z)^{-1}=\sinh(t),
\end{equation}
where $z$ is the redshift. Using eq. (\ref{eq19}) in (\ref{eq18}), we obtain the EoS parameter in terms of redshift as
\begin{equation}
\label{eq20} \omega^{de}=-\frac{1}{3}\left(\frac{1+\frac{2}{1+(1+z)^{2}}+K\frac{(1+z)^{6}}{1+(1+z)^{2}}}{1+K\frac{(1+z)^{6}}{1+(1+z)^{2}}
-\Omega^{m}_{0}(1+z)^{-3(1+\omega^{m})}}\right),
\end{equation}
\begin{figure}
\centering
\includegraphics[width=10cm,height=10cm,angle=0]{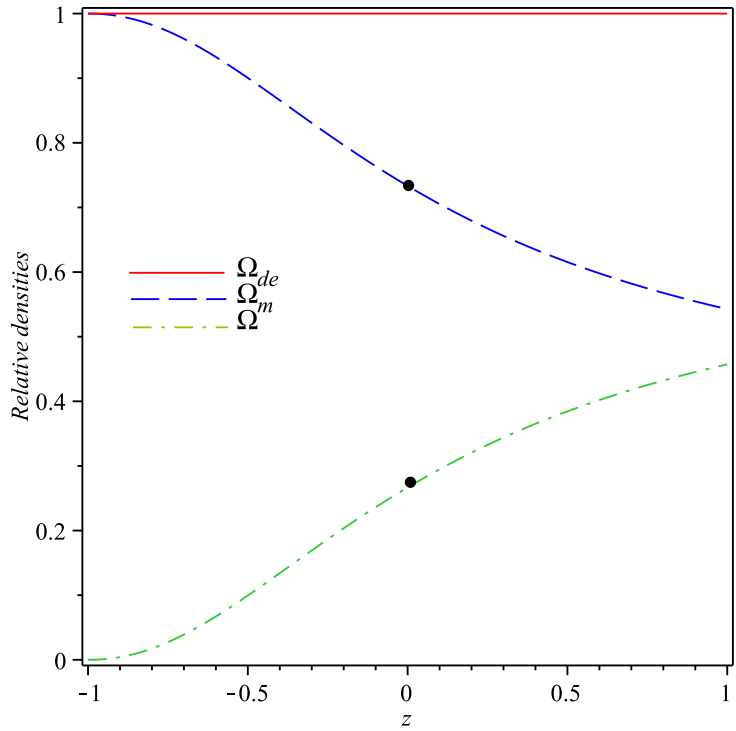}
\caption{Demo1: The plot of the DM, DE, and total energy densities ($\Omega^{m}, \Omega^{de}, \Omega$) versus $z$ for $K=0.09$. The dots show the present value of
$\Omega^{m}$ and $\Omega^{de}$.}
\label{Fig2}
\end{figure}
The behavior of EoS in term of redshift $z$ is shown in Figure 1 for different values of the anisotropy parameter $K$. It is observed that for small values of $K$, the EoS parameter is varies in quintessence region whereas for bigger values of $K$ it varies in phantom region. At the later stage of evolution it tends to the same constant value namely cosmological constant $\omega^{de}=-1$ independent to the $K$ parameter. It is worth to mention that while the current cosmological data from SNIa (Riess et al. 2004; Astier et al. 2006). CMB (Komatsy et al. 2009; MacTavish et al. 2006) and large scale structure (SDSS) (Eisenstein et al. 2005) data rule out that $\omega^{de}\geq−1$, they mildly favor dynamically evolving DE crossing the PDL (see Zhao et al. 2007; Copeland et al. 2006) for theoretical and observational status of crossing the PDL). Thus our DE model is in good agreement with the recent well established theoretical results and the the recent observations as well.\\\\
In this case, the expressions for the matter-energy density $\Omega^{m}$ and dark-energy density $\Omega^{de}$ are given by
\begin{equation}
\label{eq21}
\Omega^{m}=\frac{\rho^{m}}{3H^2}=\frac{\rho_0(1+z)^{3\left(1+\omega^m\right)}}{3\left(1+\left(1+z\right)^2\right)},
\end{equation}
and
\begin{equation}
\label{eq22} \Omega^{de}=\frac{\rho^{de}}{3H^2}=1+\frac{K\left(1+z\right)^6-\rho_0(1+z)^{3\left(1+\omega^m\right)}}{3\left(1+\left(1+z\right)^2\right)}
\end{equation}
respectively. Hence the total energy density is given by
\begin{equation}
\label{eq23}
\Omega=\Omega^{m}+\Omega^{de}=1+\frac{K\left(1+z\right)^6}{3\left(1+\left(1+z\right)^2\right)}
\end{equation}
Figure 2 shows the permitted values of $\Omega^{m}$ and $\Omega^{de}$ in our model. The dots locate the current values of these two parameters. From this figure we observe that the predicted values of these two dark components are in good agreement with those obtained through recent observations.\\

As usual we can examine our DE models through energy conditions ( For recent review see Zhang et al. 2013). The plot of weak, dominate, and strong energy conditions is shown in figure 3. Form this figure we see that in non-interacting case
\begin{equation}
\label{eq24}
(i) \rho^{de}\geq 0,~~~~~~(ii) \rho^{de}+p^{de}\leq 0, ~~~~~~ (iii)\rho^{de}+p^{de}\leq 0.
\end{equation}
\begin{figure}
\centering
\includegraphics[width=10cm,height=10cm,angle=0]{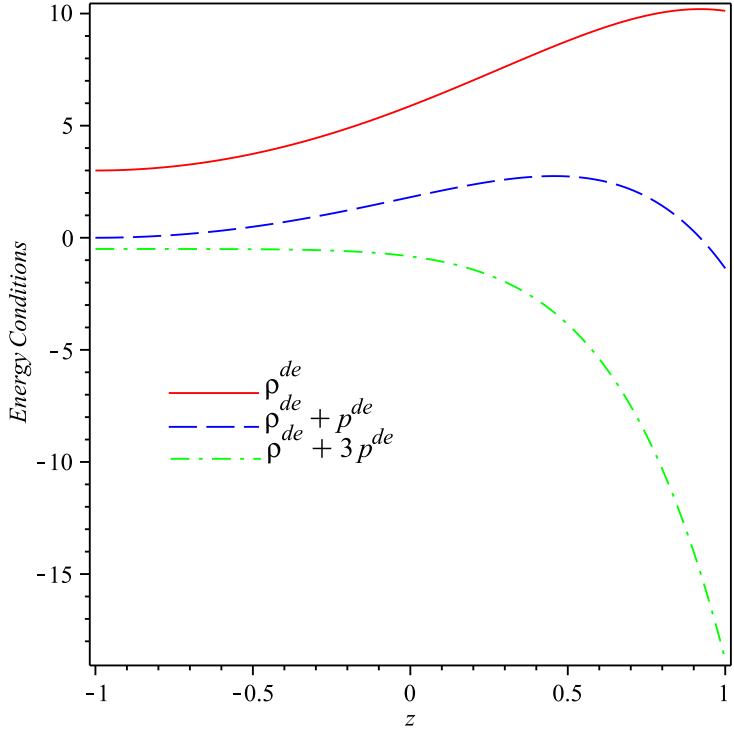}
\caption{Demo1: The plot of the weak $\rho^{de}\geq 0$, dominate $\rho^{de}+p^{de}\geq 0$, and strong $\rho^{de}+3p^{de}\geq 0$ energy conditions versus $z$.}
\label{Fig3}
\end{figure}
Thus, from the above expressions, we observe the phantom model which violates both the
strong and weak energy conditions, is a possible scenario in this case. It is worth mentioning, recent observations data indicate that the phantom model of the universe with $\omega^{de}\leq-1$ is allowed at $68\%$ confidence level.
\section{Interacting Dark Energy}
\label{sect:Interacting Dark Energy}

In this case we consider an energy transfer from DE to DM. Therefore, the interaction between the two dark components which is shown by the quantity $Q$ should be a positive function of time or equivalently redshift (see eqs.(\ref{eq25}), (\ref{eq26})). A positive $Q$ ensures that the second law of thermodynamics is fulfilled (Pavon \& Wang 2009). Here, the law of energy-conservation equation eq. (\ref{eq19}) may be written as
\begin{equation}
\label{eq25}\dot{\rho}^{m} + 3\frac{\dot{a}}{a}(1 + \omega^{m})\rho^{m}=Q,
\end{equation}
\begin{equation}
\label{eq26}\dot{\rho}^{de}+ 3\frac{\dot{a}}{a}(1 + \omega^{de})\rho^{de}=-Q.
\end{equation}
Since the nature of the dark energy still is unknown to us, we have freedom to choose different bout appropriate functions for $Q$. The most important forms of $Q$ are: (i) $Q\propto H \rho^{X}$ and (ii) $Q\propto H (\rho^{m}+\rho^{X})$. In our study we assume
\begin{equation}
\label{eq27}Q =3 H \sigma \rho^{m},
\end{equation}
where $\sigma$ is a coupling constant. Recent astrophysical observations (Guo et al. 2007) show that in the constant coupling models, $-0.08 < \sigma < 0.03$ ($95\%$ C.L.).\\

Putting eq. (\ref{eq27}) in eq. (\ref{eq25}) and after integrating, we obtain
\begin{equation}
\label{eq28}\rho^{m} = \rho_{0}^{m} a^{-3(1+\omega^{m}-\sigma)},
\end{equation}
where $\rho_{0}^{m}$ is an integrating constant. Substituting eq. (\ref{eq28}) in eq. (\ref{eq12}), we obtain
\begin{equation}
\label{eq29}\rho^{de} = 3H^{2}(1-\Omega^{m}_{0}a^{-3(1+\omega^{m}-\sigma)})-3Ka^{-6},
\end{equation}
\begin{figure}
\centering
\includegraphics[width=10cm,height=10cm,angle=0]{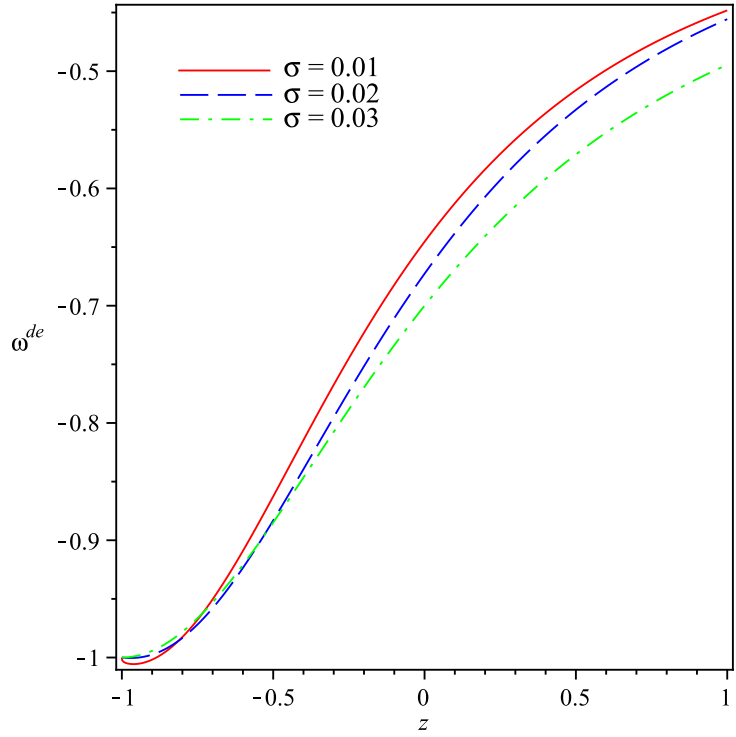}
\caption{Demo1: The plot of the dark energy EoS parameter $\omega^{de}$ versus $z$. Here, we fix the parameter $K=0.01$ and vary $\sigma$ as $0.01$, $0.02$, $0.03$.}
\label{Fig4}
\end{figure}
Using eqs. (\ref{eq10}), (\ref{eq11}), (\ref{eq28}), and (\ref{eq29}) in eq. (\ref{eq5}), the general form of the dark energy EoS parameter is obtained as
\begin{equation}
\label{eq30}\omega^{de} = \frac{2H^{2}(q-1/2)+Ka^{-6}}{3H^{2}(1-\Omega^{m}_{0}a^{-3(1+\omega^{m}-\sigma)})-3Ka^{-6}}.
\end{equation}
Using the scale factor (\ref{eq19}), we can re-write eq. (\ref{eq30}) in terms of redshift as bellow
\begin{equation}
\label{eq31} \omega^{de}=-\frac{1}{3}\left(\frac{1+\frac{2}{1+(1+z)^{2}}+K\frac{(1+z)^{6}}{1+(1+z)^{2}}}{1+K\frac{(1+z)^{6}}{1+(1+z)^{2}}
-\Omega^{m}_{0}(1+z)^{-3(1+\omega^{m}-\sigma)}}\right),
\end{equation}
The variation of the EoS parameter for dark energy in terms of red shift $z$ is shown in Figure 4. As the late time evolution of DE is interested for us, we assume $\omega^{m}=0$. In figure 4 we fix the parameter $K=0.01$ and vary $\sigma$ as $0.01$, $0.02$, $0.03$. The plot shows that the evolution of $\omega^{de}$ depends on the parameters $\sigma$ bout at the present time the dark energy EoS parameter dos not cross phantom divided line (PDL) for any value of $\sigma$. In summary, the EoS parameter only varies in quintessence region and ultimately tends to the cosmological constant region $\omega^{de}=-1$. However, the figure shows that for $\sigma=0.01$, at late time, the dark energy EoS parameter could jump to the phantom region temporary. As already mentioned, the current SNIa , CMB , and SDSS cosmological data mildly favor a dynamically evolving DE crossing the PDL.\\
\begin{figure}
\centering
\includegraphics[width=10cm,height=10cm,angle=0]{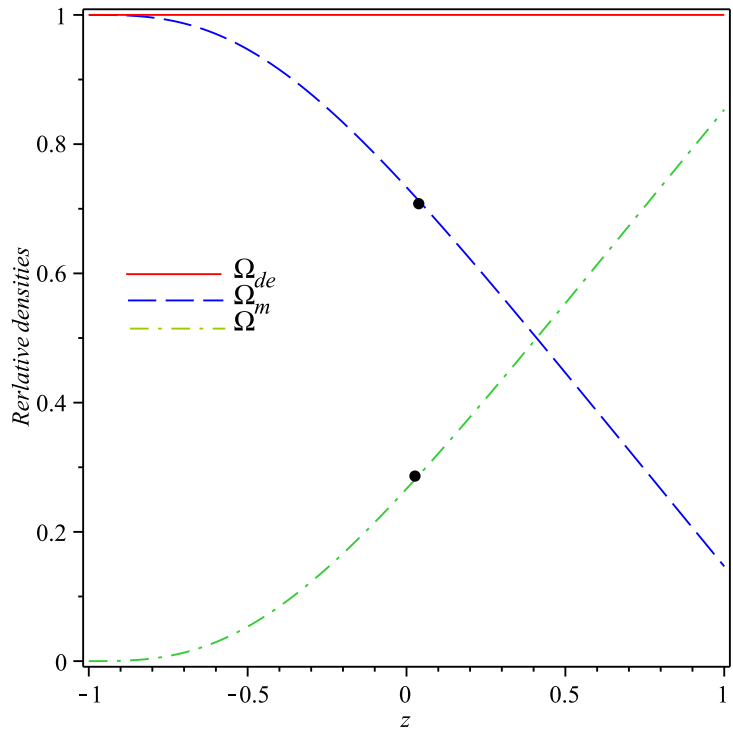}
\caption{Demo1: The plot of the DM, DE, and total energy densities ($\Omega^{m}, \Omega^{de}, \Omega$) versus $z$ for $K=0.11$. The dots show the present value of
$\Omega^{m}$ and $\Omega^{de}$.}
\label{Fig5}
\end{figure}
The expressions for the matter-energy density $\Omega^{m}$ and dark-energy density $\Omega^{de}$ are given by
\begin{equation}
\label{eq32}
\Omega^{m}=\frac{\rho^{m}}{3H^2}=\frac{\rho_0(1+z)^{3\left(1+\omega^m-\sigma\right)}}{3\left(1+\left(1+z\right)^2\right)},
\end{equation}
and
\begin{equation}
\label{eq33} \Omega^{de}=\frac{\rho^{de}}{3H^2}=1+\frac{K\left(1+z\right)^6-\rho_0(1+z)^{3\left(1+\omega^m-\sigma\right)}}{3\left(1+\left(1+z\right)^2\right)},
\end{equation}
respectively. Therefore, the total energy density is given by
\begin{equation}
\label{eq34}
\Omega=\Omega^{m}+\Omega^{de}=1+\frac{K\left(1+z\right)^6}{3\left(1+\left(1+z\right)^2\right)},
\end{equation}
which is the same as eq.(\ref{eq23}) in no-interacting case as expected. \\\\
The permitted values of $\Omega^{m}$and $\Omega^{de}$ for the interacting case are shown in figure 5. From this figure we observe that the current values of $\Omega^{m}$ and $\Omega^{de}$ predicted by our model are in good agreement with those obtained by the recent observations. In figure 5, The present values of the dark energy and dark matter energy densities are indicated by dots.\\

Figure 6 shows the plot of weak, dominate, and strong energy conditions. In this case, the energy conditions obey the following restrictions
\begin{equation}
\label{eq35}
(i) \rho^{de}\geq 0,~~~~~~(ii) \rho^{de}+p^{de}\geq 0, ~~~~~~ (iii)\rho^{de}+p^{de}\leq 0 ~~\mbox{only for}~~ \sigma>0.01.
\end{equation}
From the above expressions and figure 6, we see that in interacting case only the strong energy condition violates. Hence, in this case, the only possible scenario at the present time is quintessence.
\begin{figure}
\centering
\includegraphics[width=10cm,height=10cm,angle=0]{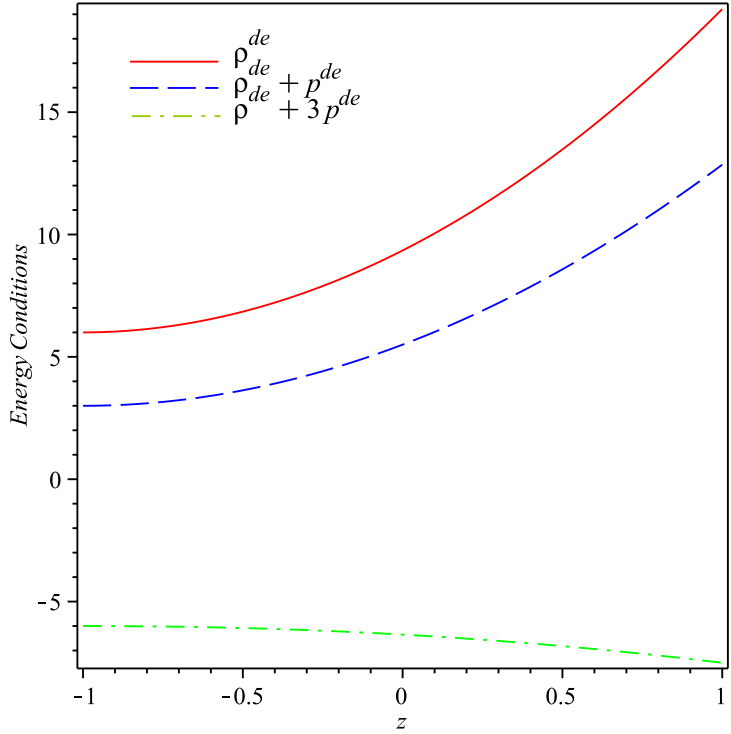}
\caption{Demo1: The plot of the weak $\rho^{de}\geq 0$, dominate $\rho^{de}+p^{de}\geq 0$, and strong $\rho^{de}+3p^{de}\geq 0$ energy  conditions versus $z$.}
\label{Fig6}
\end{figure}

\section{The Experimental $H(z)$ Test}
\label{sect:The Experimental $H(z)$ Test}
As the anisotropy parameter $K$ plays a very significant role in our study, in this section we try to fix the cosmological bound on it
by using a direct fitting procedure involving the Hubble rate $H(z)$. Here we use the so called ``differential age" method proposed by Jimenez et al
(2003) and Simon et al (2005). Later on, this method has been widely used by others to put constraints on the cosmological parameters ( for example see
Zhang et al. 2012; Zhang et al. 2011; Ma \& Zhang 2011; Luongo 2011).\\

First of all we note that in our study the scaled Hubbls  parameter is given by
\begin{equation}
\label{eq36}
E(z)=\frac{H(z)}{H_{0}}=\left[\Omega^{m}(1+z)^{2}+\Omega^{de}\frac{\rho^{de}(z)}{\rho^{de}(0)}\right]^{\frac{1}{2}},
\end{equation}
where $\rho^{X}(z)$, $\Omega^{m}$, and $\Omega^{de}$ for non-interacting case are given by eqs. (\ref{eq16}), (\ref{eq21}) and (\ref{eq22}) and for interacting case are given by
eqs. (\ref{eq28}), (\ref{eq32}) and (\ref{eq33}).\\

To constrain the model parameter $K$, we try to minimize the following reduced $\chi^{2}$.
\begin{equation}
\label{eq37} \chi^{2}_{Hub}=\sum_{i=1}^{N}\frac{[H^{th}(z_{i})-H^{obs}(z_{i})]^{2}}{\sigma^{2}_{obs}(z_{i})},
\end{equation}
where $H^{obs}$ are the values of Table. $1$, $H^{th}$ and $H^{obs}$ are referred to the theoretical and observational values for Hubble parameter respectively and the sum is taken over the cosmological dataset.
\begin{table}[ht]
\caption{The cosmological data at 1$\sigma$ error for $H(z)$ expressed in $s^{-1}MPc^{-1}Km$.
 References: 1.Simon et al. (2005) 2.Stern et al. (2010) 3.Moresco et al. (2012) 4.Gazta\~{n}age et al. (2009)  5.Zhang et al. (2012).	
}
\centering
\begin{tabular}{|c| c| c |c|}
\hline\hline{\smallskip}
$z$ & $H(z)$ & $1\sigma~error$ & Reference\\[0.5ex]
\hline
0.090 & 69 & $\pm$12 & 1\\
0.170 & 83 & $\pm$8 & 1\\
0.270 & 77 & $\pm$14 & 1\\
0.400 & 95 & $\pm$17 & 1\\
0.900 & 117 & $\pm$23 & 1\\
1.300 & 168 & $\pm$17 & 1\\
1.430 & 177 & $\pm$18 & 1\\
1.530 & 140 & $\pm$14 & 1\\
1.750 & 202 & $\pm$40 & 1\\
0.480 & 97 & $\pm$62 & 2\\
0.880 & 90 & $\pm$40 & 2\\
0.179 & 75 & $\pm$4 & 3\\
0.199 & 75 & $\pm$5 & 3\\
0.352 & 83 & $\pm$14 & 3\\
0.593 & 104 & $\pm$13 & 3\\
0.680 & 92 & $\pm$8 & 3\\
0.781 & 105 & $\pm$12 & 3\\
0.875 & 125 & $\pm$17 & 3\\
1.037 & 154 & $\pm$20 & 3\\
0.24  & 79.69 & $\pm$3.32 & 4\\
0.43  & 86.45 & $\pm$3.27 & 4\\
0.07  & 69.0 & $\pm$19.6 & 5\\
0.12  & 68.6 & $\pm$26.2 & 5\\
0.20  & 72.9 & $\pm$29.6 & 5\\
0.28  & 88.8 & $\pm$36.6 & 5\\[1ex]
\hline
\end{tabular}
\end{table}
Since we are interested in the present value of the anisotropy parameter $K$, we fix all other parameters as follows: $\Omega^{m}_{0}=0.24$, $\Omega^{de}_{0}=0.71$, $\omega^{m}=0$, $H_{0}=71$, and $\sigma=0.03$. Our results for non-interacting and interacting cases are given in Table. $2$.\\
\begin{table}[ht]
\caption{The best fit parameter with 1$\sigma$ error in non-interacting case.}
\centering
\begin{tabular}{|c|c|c|c|c|c|}
\hline\hline{\smallskip}
Case of Study & $H_{0}$ ($Km/s/M_{PC}$) & $\Omega^{m}_{0}$ &$\Omega^{de}_{0}$& $\sigma$ &$K$\\[0.5ex]
\hline
Non-Interacting Case & 71 & 0.24& 0.71& 0& 0.09 \\
\hline
Interacting Case & 71 & 0.24& 0.71& 0.03& 0.11  \\[1ex]
\hline
\end{tabular}
\label{table:nonlin}
\end{table}
\section{Conclusions}
\label{sect:conclusion}
Non-interacting and interacting dark energy with dark matter have been investigated in the scope of anisotropic Bianchi type I space-time.
In both cases, first the general form of the dark energy equation of state parameter EoS has been derived. Then we examined our general results for the case when the universe scale factor behaves as a hyperbolic function of time or redshift. It is shown that in non-interacting case, depending on the value of the anisotropy parameter $K$, the dark energy EoS parameter is varying from phantom to quintessence whereas in interacting case EoS parameter vary in quintessence region. However, in both cases the dark energy EoS parameter $\omega^{de}$, ultimately (i. e at $z=-1$) tends to the cosmological constant ($\omega^{de}=-1$). It is worth to mention that in both cases, the phantom phase is an temporary state. Carroll et al. (2003) have already mentioned that, any phantom model with $\omega<-1$ should decay to cosmological constant model $\omega^{de}=-1$ at late time. Finally, chi-squared statistical method has been used in order to constraint the model parameter $K$ with the observational data for Hubble parameter. In this case, we observed that the value of anisotropy parameter in interacting case is greater than it's value in non-interacting case. Considering the fact that at present time our universe is almost flat (i. e $K\sim 0$), above result indicates that the amount of interaction between DE and DM should be decreased as universe is expanding (or as time is going on)\footnote{One can also conclude that the space-time is much curved in interacting case.}.

\section*{Acknowledgments}
H. Amirhashchi thanks a research fund from the Mahshahr Branch of Islamic Azad University under the project entitled ``Interacting Viscous Dark Energy And Cold Dark Matter In An Anisotropic Universe``. Authors also would like to acknowledge the anonymous referee for his fruitful comments and suggestions.

\label{lastpage}

\end{document}